# Matched Filter Detection with Dynamic Threshold for Cognitive Radio Networks

Fatima Salahdine[1, 2], Hassan El Ghazi[1], Naima Kaabouch[2], Wassim Fassi Fihri[1]

[1]STRS Lab, National Institute of Posts & Telecommunication, Rabat, Morocco
[2]Electrical Engineering Department, University of North Dakota, Grand Forks, USA
Email: salahdine@inpt.ac.ma, elghazi@inpt.ac.ma, naima.kaabouch@engr.und.edu, wassim.fassifihri@gmail.com

*Abstract*—In cognitive radio networks, spectrum sensing aims to detect the unused spectrum channels in order to use the radio spectrum more efficiently. Various methods have been proposed in the past, such as energy, feature detection, and matched filter. These methods are characterized by a sensing threshold, which plays an important role in the sensing performance. Most of the existing techniques used a static threshold. However, the noise is random, and, thus the threshold should be dynamic. In this paper, we suggest an approach with an estimated and dynamic sensing threshold to increase the efficiency of the sensing detection. The matched filter method with dynamic threshold is simulated and its results are compared to those of other existing techniques.

## I. INTRODUCTION

The use of smart devices has exponentially increased over the last decade, which resulted in an increase of spectrum channels demand. However, radio spectrum is a finite resource [1]. Some channels are highly used while other are sparsely used which caused a great spectrum scarcity. Studies have shown that there is significant scope of improving spectrum utilization [2]. One of the solutions proposed to address the spectrum scarcity problem is cognitive radio. This technology allows unlicensed users, secondary users (SU), to use the radio spectrum channels allocated to licensed users, primary users (PU), when the channels are temporally not being utilized by the PUs. According to Mitola [3], cognitive radio is an intelligent radio frequency transmitter/ receiver designed to detect available channels in a wireless spectrum and change transmission parameters enabling more communications and improving radio operating behavior.

One of the important functions in a cognitive radio is spectrum sensing, which allows an SU to detect the presence or absence of PU transmission in licensed frequency bands, and allow the SU to borrow unused spectrum from the PU. A number of sensing techniques have been proposed over the last decade [4]. These techniques are classified into two categories: cooperative sensing and non-cooperative sensing [5, 6]. The non-cooperative sensing category is sub-divided into three subcategories: energy, feature, and matched filter based sensing. Energy detection is the simplest technique which does not require any information about the PU signal to operate. It works by comparing the received signal energy with an estimated threshold which depends on the noise power [7]. Correlation based sensing is a method based on the value of the autocorrelation coefficient of the received signal and exploits the autocorrelation features that exist in the transmitted signal and that are not present in the noise [8, 5]. Matched filter based sensing is a coherent pilot sensor which detects known PU signals. It maximizes the signal to noise ratio (*SNR*) at the output of the detector, but it requires a prior knowledge about the PU signal. Thus, it is assumed that the PU transmitter sends a pilot stream simultaneously with the data and the SU receiver has a perfect knowledge of the pilot stream to verify its transmission at the frequency band [1].

Sensing threshold is an important parameter in spectrum sensing. When a detector does not properly adjust its threshold, the sensing performance is degraded. A number of approaches, based on energy detection, were proposed [11-16]. As the sensing performance is highly affected by the estimation error of noise power, a dynamic estimation style of noise power is recommended in [11]. Adaptive threshold control is implemented in [12] with linear adaption of the threshold based on signal to interference plus noise ratio (*SINR*). This approach attains a considerably higher SU throughput than the fixed threshold approach, but maintains unacceptable false alarms. Adaptive threshold in unknown white Gaussian noise is presented in [13] with noise power estimation, keeping the false alarm rate at a preferred point under any noise level. It is based on a concept of a dedicated noise estimation channel in which only noise is received by SU. An improved energy detection method is proposed in [14], where misdetection of PU transmission due to sudden drop in PU transmission power is addressed by keeping an additional updated list of the latest fixed number of sensing events that are used to calculate an average test statistic value. A double-threshold technique is proposed in [15] with the aim of finding and localizing narrowband signals. Another technique is presented in [16] based on wideband spectrum sensing, which senses the signal strength levels within several frequency ranges to improve the opportunistic throughput of SU and decrease the interference to the PU.

In [7], each pair of ($P_d$, $P_f$) was associated with a particular threshold to make sensing decision. In [17], the sensing threshold is determined dynamically by multiplying the theoretical threshold by a positive factor. Several papers do not mention how the threshold was selected. However, with a static threshold, the sensing decision is not reliable because of the uncertainty of the noise. In this work, we suggest an estimated and dynamic sensing threshold for energy detection, matched filter detection and correlation based detection to increase the probability of detection and the decision reliability.

To evaluate the performance of spectrum sensing techniques, several metrics were used, including the probability of detection, $P_d$, and the probability of false alarm, $P_f$. $P_d$ is the probability that the SU declares the presence of a PU signal. The higher the $P_d$, the better the PU protection is. $P_f$ is the probability that SU declares the presence of PU when the spectrum is idle. The lower the $P_f$, the more the sensing method is effective in detecting the signal [4].

The remaining of this paper is organized as follows. In section II, we describe the sensing methods. In section III we explain the simulation methodologies for each method. The results of this simulation are discussed in section IV. Finally, the conclusion is given in section V.

## II. SPECTRUM SENSING TECHNIQUES

Spectrum sensing allows the cognitive radio users to learn about the radio environment by detecting the presence of an event using one or multiple sensors. It consists in detecting the PU signal transmission in a given time to make decision to transmit in a frequency band [7]. The spectrum sensing model can be formulated as follows

$$\begin{cases} y(n) = h * s(n) & H_0: \text{PU absent} \\ y(n) = h * s(n) + w(n) & H_1: \text{PU present} \end{cases} \quad (1)$$

where $n=1….N$, $N$ is the sample number, $y(n)$ is the SU received signal, $s(n)$ is the PU signal, $w(n)$ is the additive white Gaussian noise (AWGN) with zero mean and variance $\delta_w^2$, $h$ is the complex channel gain of the sensing channel, $H_0$ denotes the PU signal is absent, and $H_1$ denotes the PU signal is present. The output $T$ of the detector is compared to a threshold to make the right decision

if $T \geq$ threshold     then PU signal is present
if $T <$ threshold     then PU signal is absent

If PU signal is absent, SU can start to transmit its streams; if not, SU does not transmit or stops its transmission. Fig. 1 presents the general model of spectrum detection.

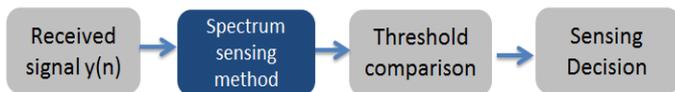

Fig. 1. General model of spectrum sensing.

This model was used for all techniques; for each technique we changed the spectrum sensing method block.

### A. Energy detection method

The decision statistic of an energy detector can be calculated from the squared magnitude of the FFT averaged over $N$ samples. The output is the received signal energy called the test statistic $T_{ED}$ as given by

$$T_{ED} = \sum_N y(n)^2 \quad (2)$$

where $n=1….N$, $N$ is the sample number, $y(n)$ is the SU received signal.

The two performance metrics ($P_d$, $P_f$) can be evaluated using the central limit theorem when $N$ is larger than 250 ($N > 250$) to approximate the test statistics as a Gaussian. These two probabilities are given by

$$P_d = Q\left(\frac{\acute{\lambda} - N(1+\gamma)}{\sqrt{2N(1+\gamma)^2}}\right) \quad (3)$$

$$P_f = Q\left(\frac{\lambda - N\delta_w^2}{\sqrt{2N\delta_w^4}}\right) \quad (4)$$

where $Q$ is the Q-function, $\lambda$ is the sensing threshold, $\gamma$ is the SNR, $\acute{\lambda}$ is the average threshold $\acute{\lambda} = \lambda/\delta_w^2$.

The sensing threshold depends on noise power. It is expressed for a target $P_f$ as

$$\lambda = (Q^{-1}(Pf)\sqrt{2N} + N)\delta_w^2 \quad (5)$$

Thus, the energy detection technique is sensitive to the noise uncertainty; however, its implementation is very simple [7].

### B. Matched filter detection with an estimated threshold

It is the optimal filter that projects the received signal in the direction of the pilot $x_p$ [10]. The test statistic is given by

$$T_{MFD} = \sum_N y(n) \, x^*_p(n) \quad (6)$$

The test statistics $T_{MFD}$ is compared with a particular threshold to give the decision. $T_{MFD}$ is a Gaussian random variable and a linear combination of Gaussian random variables. According to the Neyman-Pearson criteria [7], $P_d$ and $P_f$ are expressed as

$$Pd = Q\left(\frac{\lambda - E}{\sqrt{E\delta_w^2}}\right) \quad (7)$$

$$P_f = Q\left(\frac{\lambda}{\sqrt{E\delta_w^2}}\right) \quad (8)$$

where $E$ is the PU signal energy. Sensing threshold is giving as a function of PU signal energy and noise variance

$$\lambda = Q^{-1}(Pf)\sqrt{E\delta_w^2} \quad (9)$$

Assuming that the signal is completely known as unreasonable and impractical, some communication systems contain pilot stream or synchronization codes for channel estimation and frequency band sensing. A hybrid matched filter structure is proposed in [18] based on traditional matched filter by mixing the segmented matched filter and the parallel matched filter to overcome the frequency offset sensitivity. This structure permits the balance between sensing time and hardware complexity. As both carrier frequency offset (CFC) and phase noise (NP) demean the sensing performance of matched filter detection, in [19] matched filter detection performance is examined in presence of CFC and PN, and robust sensing techniques are proposed to overcome the negative impact of CFC and NP on detection performance.

*C. Autocorrelation based sensing*

In signal processing, for a given signal $s(t)$, the autocorrelation function is defined as [9]

$$R_{s,s}(\tau) = \int_{-\infty}^{+\infty} s(t) * s^*(t-\tau)\, dt \qquad (10)$$

The sensing decision is based on the knowledge of the statistical distribution of the autocorrelation function. For random noise, the first lag of the autocorrelation is very small or negative, but when there is a signal, the autocorrelation at the first lag represents a significant value. Thus the sensing method consists of comparing lag0 and lag1 of the autocorrelation if the received signal and sensing decision is made as following:

if *lag0* >> *lag1*    PU transmission is absent
if *lag0* ≈ *lag1*    PU transmission is present

Correlation threshold is the margin between the two lag values, for example, *lag0* is superior to ($\lambda$ %) of *lag1*, where $\lambda$ is a percentage value that represents a decision margin [3].

### III. SIMULATION METHODOLOGY

In our work, three detection methods are implemented following a specific methodology for each one. The general simulation model is presented in Fig. 2. PU signal is generated as a QPSK signal with $N$ samples. The AWGN channel adds white Gaussian noise to the input signal with the same number of samples $N$. This random noise is generated according to the *SNR* range value and the input signal power. After the spectrum sensing method is applied to the output $y(n)$, the test statistic is calculated according to each sensing method and compared with the estimated dynamic threshold. Based on this comparison, the sensing decision is made.

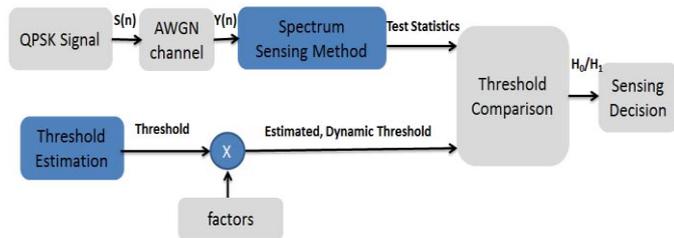

Fig. 2. Simulation model of spectrum sensing methods.

For the threshold estimation block, the threshold is estimated dynamically at each iteration, and multiplied by a threshold factor to see the impact of the threshold on detection performance. For each method, the threshold is estimated and then generated dynamically to perform the test comparison. An estimated dynamic threshold $\lambda'$ is also used to simulate $P_d$ and $P_f$ for several values of *SNR* and sample number, which can be a strong point for our simulation results. $\lambda'$ is giving as

$$\lambda' = k * \lambda \qquad (11)$$

where $\lambda$ is the estimated threshold based on each sensing method algorithm and $k$ is a positive factor.

$P_d$ and $P_f$ are used to evaluate the performance of each sensing technique. The simulation process is done for a number of iterations which is called cycle number and it represents the total number of experiments. $P_d$ is simulated as the ratio of the total number of detections, $N_d$, by the total number of experiments $N_t$. $P_f$ is simulated as the ratio of the total number of times the signal is not detected, $N_f$, by the total number of experiments $N_t$. These two probabilities are given by

$$P_d = N_d/N_t \qquad (12)$$

$$P_f = N_f/N_t \qquad (13)$$

To evaluate the performance of a given technique, the simulation process, shown in Fig. 3, is implemented and simulated for $N_t$ experiments to get $P_d$ and $P_f$. At each iteration and for each sensing parameter, a variable count which is initialized to zero is incremented by one at each iteration. If the test statistic is greater than the sensing threshold, a variable $n$ is incremented by one, and if not, a variable $m$ initialized by zero will be incremented by one. At the end of the loop, after $N_t$ experiments, the total number of detections $N_d$ is the total number of $n$ values and $N_f$ is the total number of $m$ values, so the probability of detection $P_d$ and probability of false alarm $P_f$ are the average value over $N_t$ experiments.

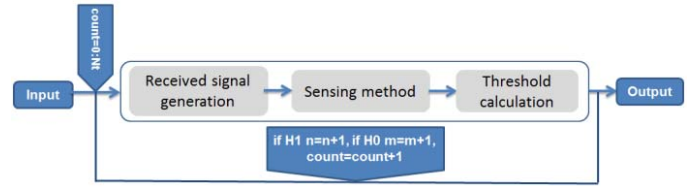

Fig. 3. Simulation algorithm for spectrum sensing methods.

The model inputs are

- Sample number is chosen as the number of samples that gives maximum value of $P_d$.
- Total number of experiments: 1000 cycles.
- SNR range: -20dB to +20dB.
- Threshold factors: 1, 2, 3, 4.

For each parameter setting, the outputs, $N_d$ and $N_f$, are obtained by varying *SNR*, or threshold, or number of samples. Afterwards, $P_d$ and $P_f$ are calculated for the different *SNR* values, $N$, and threshold.

In this paper, we chose to present the matched filter detection, detailed methodology and simulation using an estimated dynamic threshold approach. At each iteration of the loop, matched filter detection is done by convolving the received signal with some PU pilot stream, the output is averaged over $N$ samples to get the test statistics $T$ as shown in Fig.4.

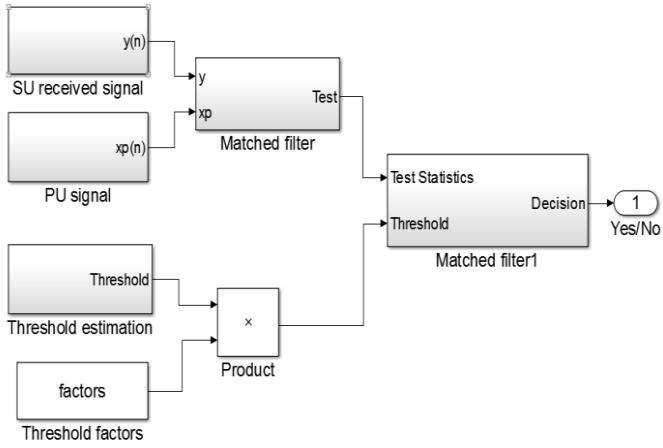

Fig. 4. Simulation model for matched filter detection.

To estimate the sensing threshold, we used the quite time approach which refers to the time period when it is assumed that PU signal is absent and only noise is transmitted. Sensing threshold is given by

$$\lambda = \sum_N w(n) x_p^*(n) \qquad (14)$$

where $\lambda$ represents the estimated threshold value, $n=1\ldots N$, $N$ is the sample number, $w(n)$ is the additive white Gaussian noise, $x_p^*(n)$ is the PU pilot stream. At each iteration, the threshold value is generated dynamically using the quit time approach and multiplied by a threshold factor (1 to 4) to investigate the impact of the threshold on the performance of the sensing technique.

## IV. RESULTS AND DISCUSSION

The results of the performance of the matched filter were compared to those of energy detection and autocorrelation techniques. To compare the three methods, simulations were done using the same parameters and varying *SNR*, threshold and number of samples *N*. Fig. 5 shows $P_d$ as a function of *SNR* for the three aforementioned sensing techniques. In this simulation $N=1000$, and the threshold factor $=1$.

As one can see, $P_d$ increases with *SNR*. For *SNR* values higher than 0dB, $P_d$ is higher than 90%. Under high *SNR* values, in which the signal is higher than noise, matched filter detection achieves its 100% of detection with a small number of samples while energy detection presents a very high probability of detection. However, false alarm is very high for energy detection, which means that this detector is not able to distinguish between noise and signal. Autocorrelation based sensing presents a lower probability of detection comparing to the other methods. However, the lack of false alarm data does not allow us to compare the performances of the three sensing techniques.

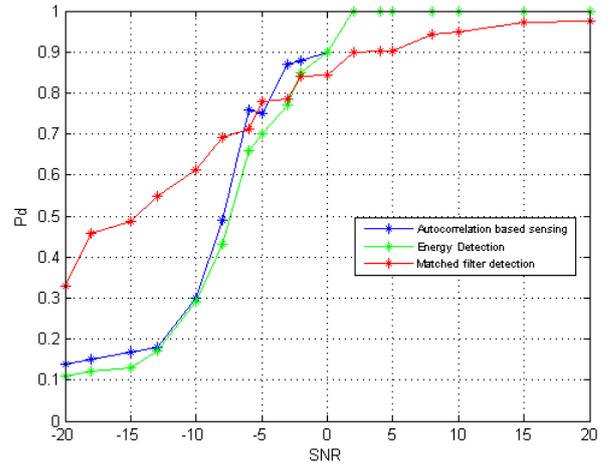

Fig. 5. Simulation results for sensing methods comparison $P_d$ vs SNR.

To summarize, in order to obtain a good detection, each method has a particular behavior under certain environment conditions. For energy detection, it gives a good performance under high *SNR* values, and, thus works best only for high power signals, high threshold, and high number of samples. However, this method is not reliable because of the non-ability to distinguish between noise and signal after detection. On the other hand, matched filter sensing achieves a high performance for low sample number and can work under low *SNR* values. For the autocorrelation based detection, the results give low $P_d$ under low *SNR* compared to the others methods.

In the second experiment, we focused on matched filter detection results. The simulation is done for 1000 experiments to get an average result of $P_d$ and $P_f$ for each set of parameters.

Fig. 6 represents $P_d$ as a function of *N* for several values of *SNR*. For higher *SNR* (>- 4dB), $P_d$ is nearly equal to 100%. For lower *SNR* (<- 20dB), $P_d$ increases for *N* greater than 400 samples. $P_d$ also increases with the increase of *N* and it achieves 100% value for $N=1000$. Thus, for small number of samples, the matched filter detector can work even at low *SNR*, and achieves 100% for 1000 samples.

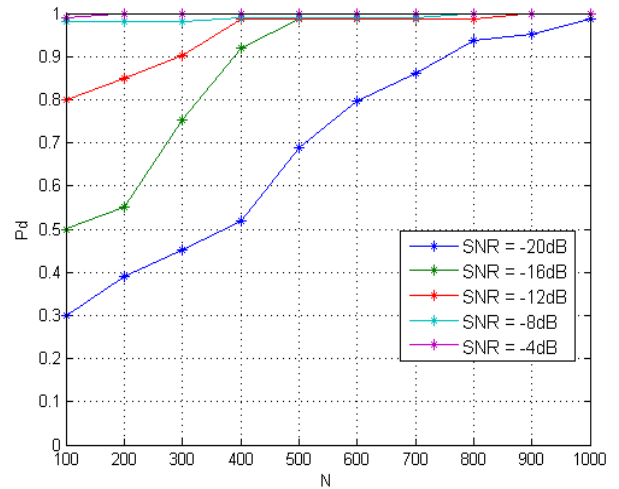

Fig. 6. Simulation results for $P_d$ vs. N with variable SNR.

$P_d$ and $P_f$ were calculated by varying the *SNR* from -20dB to +20dB for several values of threshold factor, *N* was chosen equal to 1000. Fig. 7 presents the $P_d$ as a function of *SNR* for several values of threshold factor. For fixed threshold factor, $P_d$ increases with *SNR*. As one can see, $P_d$ decreases when the threshold factor increases.

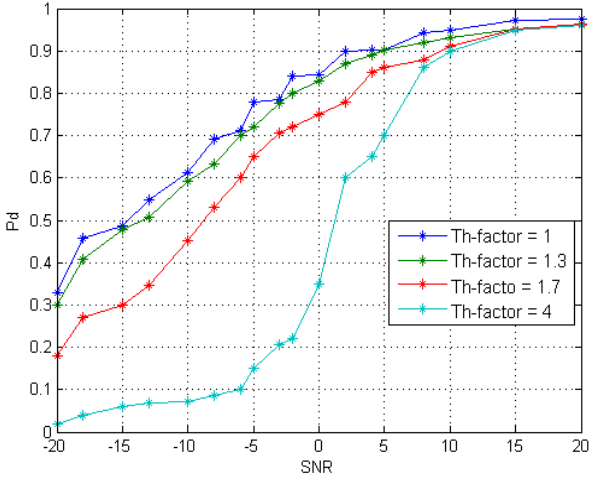

Fig. 7. Simulation results for $P_d$ vs. SNR with variable threshold.

Fig. 8 presents $P_f$ as a function of *SNR* for several values of threshold factor. For a high threshold factor (*Th_factor* = 4), $P_f$ decreases for a low *SNR* values (<- 10dB); and when the threshold factor decreases, $P_f$ increases. For a fixed threshold factor and low *SNR*, $P_f$ is high. This figure also show the impact of the threshold on $P_f$ which decreases with the increase of threshold and *SNR*.

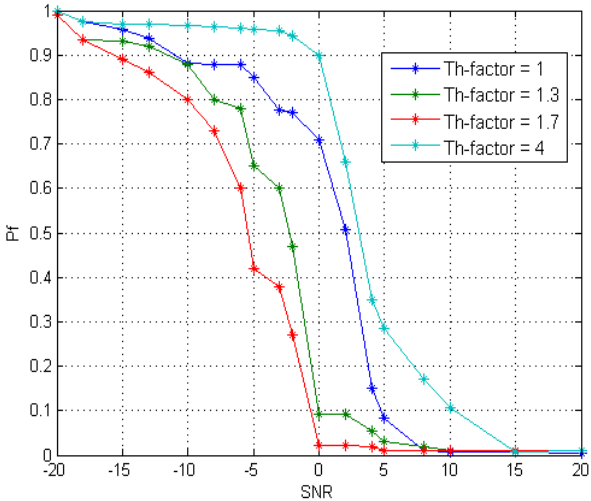

Fig. 8. Simulation results for $P_f$ vs. SNR with variable threshold.

These simulation results and others show that matched filter sensing technique can work under low *SNR* and with a small number of samples.

Table I presents a comparison between the proposed simulation and the other spectrum sensing techniques, energy detection and autocorrelation.

TABLE I. SPECTRUM SENSING TECHNIQUES COMPARISON

| **Energy detection** | **Autocorrelation based sensing** | **Matched filter detection** |
|---|---|---|
| Low complexity | High data processing | High complexity |
| Prior knowledge of PU signal is not required | Knowledge of statistical distribution of the autocorrelation function is required | Prior knowledge of PU signal is required |
| - Non-ability to distinguish between noise and signal after detection <br><br> -Good performance under high SNR | - Ability to distinguish between PU signal and noise after detection <br><br> - Robust against noise uncertainty | - High performance for low sample number <br><br> - Ability to perform under low SNR region |

## V. CONCLUSION

In this paper, we described the basic sensing methods and their simulation results were discussed. From the results of the comparison, we can conclude that each method has its strengths and weaknesses. Energy detection is easy to implement and does not require any information about PU signal; however it is not able to distinguish between signal and noise. It has also a high false alarm. Matched filter sensing requires a perfect knowledge of PU signal, which is not practical, but has a good performance under low *SNR*. Nevertheless, autocorrelation based sensing is robust against noise uncertainty. Therefore, choosing one of these three sensing techniques depends on the *SNR* level, noise uncertainty of the transmission channel, and available information about the PU signal. In addition, using a dynamic threshold gives better sensing performance than when using a static threshold.